# Measurements of complex refractive indices of photoactive yellow protein

KyeoReh Lee[1,+], Youngmin Kim[2,3,+], JaeHwang Jung[1], Hyotcherl Ihee[2,3,*], and YongKeun Park[1,*]

[1]Department of Physics, Korea Advanced Institute of Science and Technology, Daejeon 305-701, Republic of Korea.
[2]Center for Nanomaterials and Chemical Reactions, Institute for Basic Science (IBS), Daejeon 305-701, Republic of Korea.
[3]Department of Chemistry, Korea Advanced Institute of Science and Technology, Daejeon 305-701, Republic of Korea.
[*]yk.park@kaist.ac.kr and hyotcherl.ihee@kaist.ac.kr.
[+]these authors contributed equally to this work

**ABSTRACT**

A novel optical technique for measuring the complex refractive index (CRI) of photoactive proteins over the wide range of visible wavelengths is presented. Employing quantitative phase microscopy equipped with a wavelength swept source, optical fields transmitted from a solution of photoactive proteins were precisely measured, from which the CRIs of the photoactive proteins were retrieved with the Fourier light scattering technique. Using the present method, both the real and imaginary RIs of a photoactive yellow protein (PYP) solution were precisely measured over a broad wavelength range (461 – 582 nm). The internal population of the ground and excited states were switched by blue light excitation (445 nm center wavelength), and the broadband refractive index increments of each state were measured. The significant CRI deviation between in the presence and absence of the blue excitation was quantified and explained based on the Kramers-Kronig relations.

**Introduction**

Light-matter interactions constitute an important part of characterizing a sample using diverse existing methodologies ranging from visual inspection, scattering analysis, and microscopy to spectroscopy. As light is electromagnetic wave, the light-matter interaction is the composite of electric- and magnetic-response of matter that can be quantified by the single parameter called the complex refractive index (CRI),

$$\tilde{n}(\omega) = \sqrt{\left[1 + \chi_e(\omega)\right]\left[1 + \chi_m(\omega)\right]} = n(\omega) + i\kappa(\omega), \quad (1)$$

where $n$ and $\kappa$ are the real and imaginary RIs; $\chi_e$ and $\chi_m$ are the electric and magnetic susceptibilities; and $\omega$ is the temporal frequency of light; respectively. Therefore, light-matter interaction can be interpreted by the CRI, and its spatiotemporal variation.

In particular, directly measuring the CRI of proteins is of great interest to physical chemists and structural biologists trying to unravel the structure and functions of proteins because the CRI of a protein is strongly related to its structure and function. It is especially advantageous for those proteins that cannot be crystallized for X-ray analysis, or are too large for conventional NMR spectroscopy. For example, circular dichroism spectroscopy that analyzes the CRI difference between the two circular polarization of light is sensitive to the secondary structures of proteins[1].

Unfortunately, despite such powerful molecular characterization capabilities, direct measurement of the CRI of proteins over a wide range of wavelengths has been hindered by limitations in existing instrumentations. Generally, the real and imaginary RIs are independently measured with a reflectometer and absorption spectrometer, respectively, but not simultaneously. Kramers-Kronig (K-K) relations can be utilized to estimate the real RI from the measured imaginary RI or vice versa. However, the K-K relations have problems with accuracy and quantification unless the detection wavelength range is broad enough. Alternatively, spectroscopic ellipsometry can measure the CRI of a thin film[2], but in this case the CRI information can be extracted only with a model-based analysis to describe light-matter interactions. More importantly, it is challenging to measure the RI changes associated with the conformational changes of proteins caused by external stimulation such as in photoactive proteins by using existing methods.

On the other hand, quantitative phase imaging (QPI) techniques[3,4] have a potential to circumvent these limitations and offer a unique approach to measuring the CRIs of proteins. QPI techniques use the principle of holography to quantitatively and precisely measure both the optical attenuation and phase delay images, or optical field images of a sample, which is related to the CRI of the sample[3,4]. The optical field images of both non-biological and biological samples, including colloidal particles[5,6], red blood cells[7,8], neurons[9,10], and tissues[11] have been measured precisely and quantitatively by QPI techniques.

Herein, we present a novel method to measure the CRI changes of photoactive proteins over the wide range of visible wavelengths. Using the QPI technique combined with a wavelength swept source[12,13], the optical light fields transmitted from proteins were precisely measured. Then, the CRIs of proteins were retrieved by analyzing the angle-resolved light scattering spectra of the transmitted light fields with the Fourier light scattering technique[14]. Using the method, we measured the broadband CRI of a photoactive protein, photoactive yellow protein (PYP)[15-18]. Furthermore, by illuminating the proteins with a beam of excitation, RI increments of both the ground and photoactive states were quantified over the visible wavelengths. PYP is a small (14 kDa) and highly soluble protein related to phototaxis signal transduction in *Halorhodospira halophile*. The chromophore of a PYP, p-coumaric acid absorbs blue light (absorption peak at 446 nm; see Supplementary Fig. S3 online) and triggers the photocycle. The later part of the photocycle has a state called *pB* which exhibits structural change, blue-shifted light absorption, and the longest relaxation time among the intermediate states[19-22]. The real RI values of PYP has not been extensively studied and only qualitatively estimated[23], whereas the imaginary RI or absorption has been studied thoroughly. For the rest of article, we denote the ground and excited states of PYP as the *pG* and *pB* states, respectively.

### Results and Discussion

The measured CRI values of both the *pG* and *pB* states of the PYP solution are shown in Fig. 2. Both the real and imaginary RIs of the PYP were determined by fitting the Mie theory results to the measured angle-resolved light scattering patterns.

The imaginary RIs ($\kappa$) of the PYP decreases monotonically as the wavelength increases (Fig. 2a) and converges to zero for wavelengths longer than 500 nm. The $\kappa$ in the presence of the pump beam (pump−on state) are approximately five fold smaller than those of PYP in the absence of the pump beam (pump−off state). The negative mean values of $\kappa$ at 582 nm, and the increasing inaccuracy for the longer wavelength is due to the wider bandwidth of probe beam that decreases interference efficiency.

The significant decrease of $\kappa$ in the presence of the pump beam indicates a population transition from the *pG* to the *pB* state. As the absorbance of *pB* is negligible in current spectral domain[24], we deduced the non-zero $\kappa$ in pump−on state is originated from the remaining *pG* population.

Thus, from the measured $\kappa$, the molecular density of *pG* ($\rho_{pG}$) can be determined from the known molecular extinction coefficient $\varepsilon$ of *pG* as

$$\kappa(\rho,\lambda) = \frac{\ln(10)}{4\pi} \rho_{pG} \lambda_b \varepsilon(\lambda_b), \qquad (2)$$

where $\lambda_b$ is the wavelength of a probe beam. The $\rho_{pG}$ were measured as 3.44 ± 0.1 mM and 0.56 ± 0.1 mM, for the pump−off and −on states, respectively (Fig. 2a). Assuming the relaxation time of intermediate states between *pG* and *pB* states are very short and would not significantly contribute to the population equilibrium, the two states of PYP molecule are complementary. Therefore, the molecular density of *pB* state is 2.88 ± 0.1 mM, and the *pB* population ratio ($R_{pB}$) is 0.837 ± 0.035. A study by using LED for illumination showed a similar $R_{pB}$[25].

The real RIs of the PYP also decreases monotonically as the wavelength increases (Fig. 2b). The real RIs of the pump−off state decreased from 1.3518 to 1.3452 for a wavelength ranging from 461 to 582 nm. The real RIs of the pump−on state are lower than those of the pump−off state for all the wavelengths in the range. This decrease in the real RI in the presence of pump beam illumination is consistent with the decrease in the population densities of the light-absorbing *pG* state. The real RIs of the pump−on state decreased from 1.3510 to 1.3450 for a wavelength ranging from 461 to 582 nm.

From the retrieved density population of protein states, the refractive index increment (RII, $\partial n/\partial \rho$) of *pG* and *pB* state PYP solution can be calculated individually using the following equation:

$$\begin{bmatrix} n_{pump-off} \\ n_{pump-on} \end{bmatrix} = \begin{pmatrix} C_{pB,pump-off} & C_{pG,pump-off} \\ C_{pB,pump-on} & C_{pG,pump-on} \end{pmatrix} \begin{bmatrix} RII_{pB} \\ RII_{pG} \end{bmatrix} + n_0, \quad (3)$$

where $C_{pB, pump-off}$ corresponds to the concentration of *pB* states in the pump-off case, and $n_0$ is the real RI of surrounding medium. The results are shown in Fig. 2c, and tabulated in Supplementary Table. S2 online. The difference in RIIs between the *pG* and *pB* PYP solution was maximized (approx. 0.3 M$^{-1}$) at 470 nm, and decreased as the wavelength increase.

The *pB* population ratio $R_{pB}$ of a state is related to the dynamic kinetics of the transition between the *pG* and *pB* states of the PYP. The relaxation time $\tau$ of the *pB* state can be obtained from the measured $R_{pB}$ with the following equation[26,27]:

$$\frac{1}{\tau} = \ln(10) \left( \frac{1 - R_{pB}}{R_{pB}} \right) \frac{\phi}{N_A hc} \int \varepsilon(\lambda_P) \lambda_P \frac{\partial I_P}{\partial \lambda_P} d\lambda_P, \quad (4)$$

where *h* is Planck's constant; $N_A$ is Avogadro's constant; $\lambda_P$ is the wavelength of a pump light; $\phi$ is the photocycle quantum yield of PYP; amd $\partial I_P/\partial \lambda_P$ is the spectral density of the pump beam. Inserting $\phi$ = 0.35 from the literature[28,29]; and $\partial I_P/\partial \lambda_P$ (see Supplementary Fig. S2 online), and $R_{pB}$ = 0.837 ± 0.035 from the measurement, $\tau$ is calculated to be 77 ± 27 ms. We note that the relaxation time determined here is smaller than those reported by typical time-resolved pump-probe experiments (150 ms – 2 s)[20,30,31]. The discrepancy may be related with the different mode of data collection (continuous illumination vs pump-probe) but the exact origin is not clear at this stage. The accuracy of this calculated $\tau$ is mainly determined from the uncertainty of $R_{pB}$ due to the high $R_{pB}$ sensitivity of $\tau$ in Eq. (4).

To further validate the accuracy of the measurements, we used the K-K relations. The K-K relations connect the real and imaginary RIs based on the causality of the response functions. The K-K relations provide a solution for the real RI from the imaginary RI as

$$n(\omega) = 1 + P \int_{-\infty}^{+\infty} \frac{d\omega'}{\pi} \frac{\kappa(\omega')}{\omega' - \omega} \quad (5)$$

or vice versa. However, because of its inherent integral form, the K-K relations do not directly provide an exact quantitative solution. Nonetheless, the K-K relations provide qualitative trends for *n* from the measured $\kappa$.

Using the K-K relations, we obtained the trends for the real RIs from the measured imaginary RIs, and compared them with the measured real RIs. The results are shown in Fig. 3. The overall shape of the function *n*, obtained from the K-K relations, is well matched with the experimentally measured *n*, for both cases with the pump beam turned on and off, respectively. The function *n* obtained from the K-K relations has a peak at a wavelength of 470 nm, which is consistent with the measured *n*. This result further shows the reliability and accuracy of the measured CRI values of the PYP which reflect the structural changes in the PYP.

The relation between the CRI and electromagnetic susceptibilities (Eq. 1) should be emphasized, because the susceptibilities of molecules strongly relate to their electric-, and magnetic-dipole moments. Therefore, measurements of CRIs for specific molecules can provide clues for protein structures. For example, Tamasaku *et al.* visualized electron cloud distributions of diamond with resolution of 0.54 Å using extreme-ultraviolet EM wave (103 Å – 206 Å) and the non-linear susceptibility relationship between X-ray and the used frequenceis[32]. In the current study, the CRI measurements of PYP proteins were performed in solution. Thus, it is challenging to directly translate our CRI measurements into structural changes in PYP because proteins in solution have arbitrary orientations that overall smoothen the directional information. When directional-, and/or polarization-dependent CRIs or the electromagnetic susceptibilities of proteins are systematically measured, it can provide important information on the protein structures.

## Conclusions

Herein, we present a method to precisely and quantitatively measure the CRIs of photoactive proteins over a wide range of wavelengths. Using a quantitative phase microscopy equipped with a wavelength-sweeping source and Fourier transform light scattering technique, CRIs of PYP have been measured for wavelengths ranging from 461 to 582 nm.

We found a significant difference in the CRI values of PYP in the pump–off and the pump–on states, not only for the imaginary RI, but also for the real RI. More importantly, we retrieved the RII values of PYP, concentration-normalized refractive index properties of molecules, on both the *pG* and *pB* states. Our measurements have been verified with independent UV-VIS absorption spectroscopy and the K-K relations. We expect the present method can be widely used to measure the CRIs of photoactive proteins in various fields including biology, chemistry, medical science, pharmacy, and food industry.

## Methods

### Purifying PYP

The pQE80L-PYP plasmid containing the PYP gene was transformed into *E. coli* BL21 (DE3). Next a freshly grown colony from the transformed bacterial culture was used to prepare a 50 ml LB broth seed culture. The seed culture was further used to inoculate 12 liters of LB for a large culture at 37°C. When the optical density (O.D.) of the large culture reached 0.6, IPTG was added to the culture so to have the final concentration as 1 mM, and the temperature was reduced to 18°C and incubated overnight. After culturing overnight, the culture was centrifuged at 6000 rpm for 15 minutes to harvest the cells. The cells were then dissolved in a 20 mM Tris buffer (pH 7.0) containing 150 mM NaCl and 1 mM PMSF, and sonicated for cell disruption. The chromophore precursor (ρ-coumaric acid) was added the disrupted cell lysate. The mixture was centrifuged at 17000 rpm for 1 hour to separate the cell debris. The supernatant was subjected to nickel affinity chromatography and purified with the gradient elution method. The purified PYP solution was dialyzed with 20 mM Tris-Hcl, pH 7.0 buffer and concentrated. The purity of the concentrated PYP solution was increased with the Hitrap Q column for ion exchange chromatography. The final PYP solution was mixed with 100 μm diameter PMMA microspheres (74214 FLUKA, Sigma-Aldrich, Inc.) that were washed three times with distilled water. Ten microliters of the mixture was sandwiched between coverslips, and the edges were sealed with epoxy adhesive.

### Optical setup

The optical setup consists of an illumination part, in which the wavelength of the illumination to be used as the probe beam is systemically scanned over a broad visible spectral range, and an optical field measurement part that a quantitative phase imaging unit (QPIU)[33] is implemented. The QPIU is common-path full-field interferometry which uses the principle of lateral shearing interferometry (see Supplementary Fig. S1 online for the detailed optical setup ).

For each *pG* and *pB* state, we measured holograms at 11 different wavelengths ranging from 461 to 582 nm. To modulate the state population of a PYP solution, a blue LED (455 nm peak wavelength, M455L3, Thorlabs Inc.) was used with a band-pass filter (438 ± 12 nm, FF02-438/24-25, Semrock, Inc.) to provide a pump beam. Supplementary Table. S1, and Fig. S2 online show detailed specifications of used beams. Although there exists a spectral overlap between the pump beam and the probe beam of the shortest wavelength (461 nm), the pump beam does not affect the RI measurement because the temporal incoherency of the LED prevents the formation of interference patterns.

### Experimental procedure

In order to obtain the CRI of PYP, we first measured the wavelength-dependent light field images of a 100−μm−diameter PMMA microsphere immersed in PYP solution (Fig. 1). The measurements were performed with and without the pump beam (445 nm) to measure the CRI of both the *pG* and *pB* states. To ensure pure pG state of the PYP solution, we allowed a sufficient idle time (>10 seconds) between switching on/off the LED. From the measured raw holographic image (Fig. 1b), the amplitude and phase images of the bead were obtained with a field retrieval algorithm (Figs. 1c–d)[34]. From the measured light field images, the angle-resolved light scattering patterns of the microsphere were calculated with the Fourier transform light scattering technique (FTLS)[12-14] shown in Fig. 1e. In FTLS, the measured optical field of a sample is numerically propagated to the far-field with the Fraunhofer diffraction theory which is implemented with 2-D Fourier transform[13,14]. Then, the obtained angle-resolved light scattering pattern is fitted to the Mie scattering theory from which the RI of the PYP solution is precisely obtained.

### CRI extraction

To determine the CRI of the PYP solution, we performed the nonlinear fitting of the measured FTLS signals to the Mie theory. We used three fitting parameters: the real RI (*n*) and the imaginary RI (*κ*), and the diameter of the PMMA microsphere. For a robust and automated fitting analysis, the angle-resolved light scattering signals were prepared as a function of $n\sin\theta$ instead of $\theta$, in order to ensure the same scale for the horizontal axes while adjusting the *n*.

Due to the highly oscillatory features of the angle-resolved light scattering curves (Fig. 1e), the direct application of conventional fitting algorithms shows non-convergent results, or is dependent on the initial fitting parameters. Instead, we used a process prior to the fitting processes to find the appropriate initial fitting parameters. By minimizing the total variance of the extreme positions between the measurement and the Mie theory. These initial parameters yield highly reproducible fitting results. Thereafter, we used a nonlinear fitting algorithm (*nlinfit*, a built-in function of MatLab$^{TM}$) with the predetermined initial fitting parameters. In order to retrieve the CRI values of PYP solution, the CRI of the PMMA microspheres should be recalibrated because the measured FTLS signals reflects the difference in CRI between PYP solution and the PPMA microspheres. The CRI of the PMMA microspheres was calibrated using the identical method in which the PMMA microspheres were immersed in distilled water whose CRI values are known[35].

## Acknowledgements

This work was supported by KAIST, the Korean Ministry of Edu-cation, Science and Technology (MEST), and the National Research Foundation (2012R1A1A1009082, 2013K1A3A1A09076135, 2013M3C1A3063046, 20090087691, 2012-M3C1A1-048860, and IBS-R004-G2-2014-a00).

## AUTHOR CONTRIBUTIONS

Y.-K.P., and H.L. conceived and supervised the study. K.R.L and Y.K. performed the optical measurement and analyzed the data. J.H.H. built the optical setup and provided the analysis tool. All authors discussed the experimental results and wrote the manuscript.

## COMPETING FINALCIAL INTERESTS

The authors declare no competing financial interests.

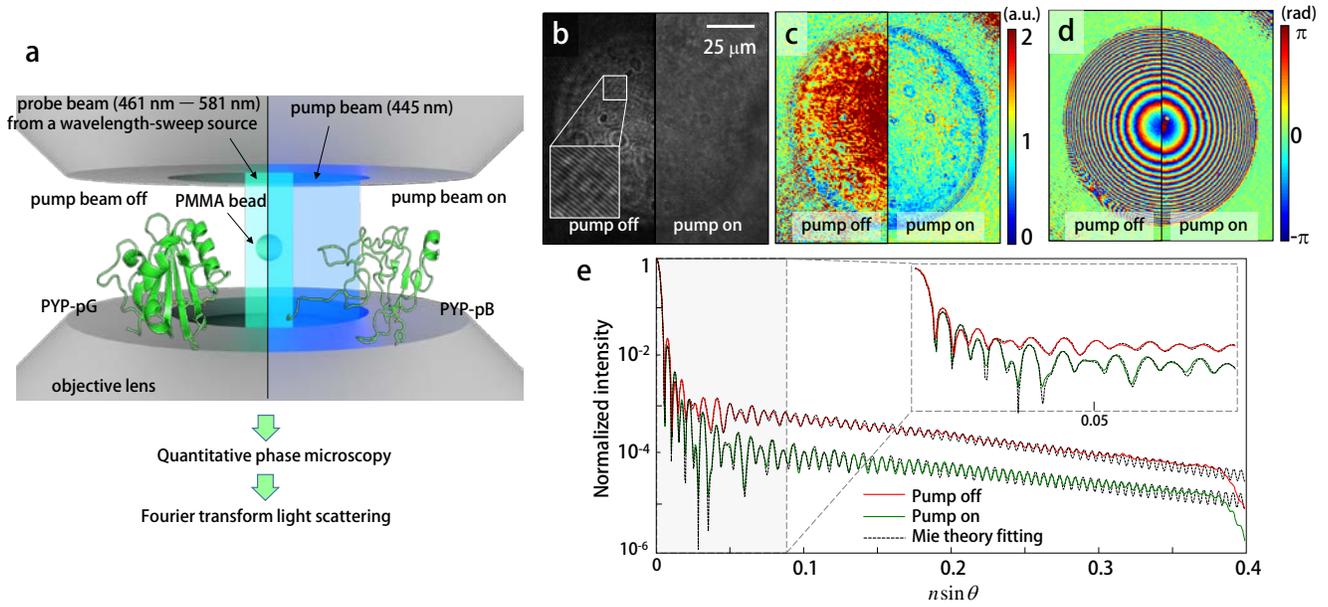

**Figure 1.** Experimental procedure for measuring the CRI of PYP protein solution. **(a)** A conceptual schematic of the measurements. The optical field of a microsphere immersed in a PYP solution is obtained over a broad range of visible wavelengths. *Left*, in the absence of a pump beam, PYP remains in a ground state (*pG*). *Right*, in the presence of a pump beam, PYP is switched to an excited state (*pB*). **(b)** Raw hologram of a PYP solution without (left) and with the pump beam (right). **(c)** Measured amplitude and **(d)** phase images of the PYP solution. **(e)** Retrieved angle-resolved light scattering spectra of the PYP solution without (red), and with the pump beam (green). The solid and dotted lines represent the experimental and theoretical results, respectively. *Inset*, scattering spectra within the smaller scattering angle range.

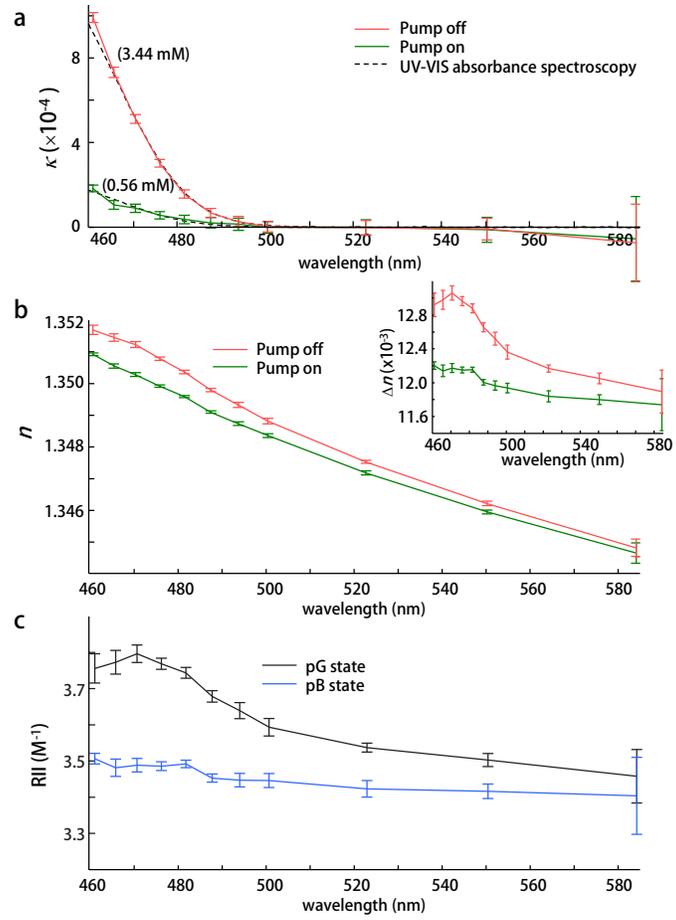

**Figure 2.** The measured CRI and RII of PYP solution. **(a)** Imaginary RI values. The black dotted lines are fitted graph based on Eq. (2). Corresponding molecular densities are indicated in brackets. **(b)** Real RI values, and *Inset*, the difference in real RI between PYP solution and distilled water in the pump–off (red) and the pump–on state (green). **(c)** RII values of *pG* (black), and *pB* (blue) states. All error bars indicate the standard deviation from 5 measurements with different microspheres.

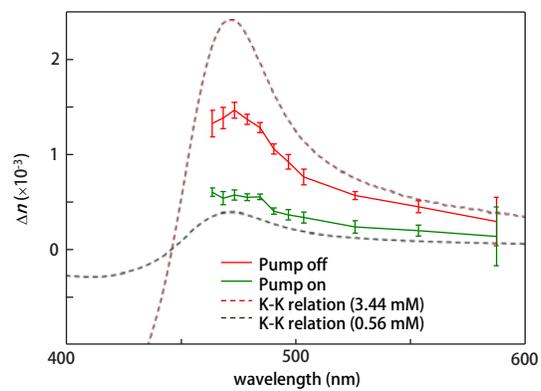

**Figure 3.** Calculated real RI trends of the PYP, based on the K-K relations (dotted lines). The real RI difference from distilled water are also shown for comparison purposes (solid lines). Please note that the *y*-axis offset of dotted lines is set discretionary because of the ambiguity from the integral forms of the K-K relations.